# Size effects in the magnetic behaviour of TbAl$_2$ milled alloys

D P Rojas, L Fernández Barquín, J Rodríguez Fernández, J I Espeso and J C Gómez Sal

Departamento CITIMAC, Universidad de Cantabria, Santander 39005, Spain

E-mail:rojasd@unican.es

**Abstract.** The study of the magnetic properties depending upon mechanical milling of the ferromagnetic polycrystalline TbAl$_2$ material is reported. The Rietveld analysis of the X-ray diffraction data reveals a decrease of the grain size down to 14 nm and -0.15 % of variation of the lattice parameter, after *300 hours* of milling time. Irreversibility in the zero field cooled - field cooled (ZFC-FC) DC-susceptibility and clear peaks in the *AC* susceptibility between 5 and 300 K show that the long-range ferromagnetic structure is inhibited in favour of a disordered spin arrangement below 45 K. This glassy behaviour is also deduced from the variation of the irreversibility transition with the field ($H^{2/3}$) and frequency. The magnetization process of the bulk TbAl$_2$ is governed by domain wall thermal activation processes. By contrast, in the milled samples, cluster-glass properties arise as a result of cooperative interactions due to the substitutional disorder. The interactions are also influenced by the nanograin structure of the milled alloys, showing a variation of coercivity with the grain size, below the crossover between the multi- and single-domain behaviours.

## 1. Introduction

The study of *heterogeneous magnetic structures* is a matter attracting a vast research effort due to the rich phenomena related to the existence of competing interactions. These may lead to the promotion of magnetic disorder and, as a consequence, to the existence of a variety of magnetic ground states commonly associated to spin-glass or reentrant spin-glass behaviours [1]. It is frequent that heterogeneous structures can also display some degree of *magnetic clustering*, as occurs for instance in colossal magnetoresistance oxides [2], strongly correlated systems [3] or amorphous Fe-Zr alloys [4,5] to cite a few examples. In addition, many heterogeneous structures are included in the *nanomagnetism* research field as they are formed by magnetic nanograins in a metallic or insulating matrices, giving rise to giant magnetoresistance, exchange-bias and single-domain magnetic properties [6-8]. It is then obvious the existing relationship between the magnetic disorder and the existence of magnetic clusters of different sizes depending upon the sample, enhancing or masking the different magnetic interactions.

The influence of disorder on the magnetic properties of the system comprising cubic GdX$_2$ ( where X = Al, Pt, Ir, Rh and Mg) Laves phase compounds has been studied recently [9-14]. The underlying interest is connected to our above statements; this is a ferromagnetic system very sensitive to disorder (caused by atomic substitutions or milling processes) and, hence suitable to observe what degree of disorder is necessary to break down the long-range ferromagnetism. In particular, it was observed that upon milling, the value of the Curie temperature ($T_C$) depends on the crystallographic change in the lattice parameter and the electronic character of the conduction electrons [11]. The ferromagnetic contribution disappears and a glassy magnetic behaviour emerges at lower temperatures. Zhou and Bakker also discussed about the intrinsic nature of the emerging phase that appears for long grinding times, when the crystallite size becomes constant [9]. Additionally, other analyses have considered the role of grain boundaries in these alloys in relation to the random anisotropy effects [12, 14]. However such a role remains relatively unexplored although it presents some similarities to other 3d transition metal-milled samples [15]

In some studies [10,13], the *AC*-susceptibility revealed a curious behaviour in which the influence of domain wall relaxation was apparent. In these studies, no particular discussion was brought afore in relation to possible size effects in the milled samples, which is surprising considering that the decrease of particle size when using mechanical alloying processes is a very common result in other metallic or non-metallic systems [16]. The *AC* susceptibility ($\chi_{ac}$) is convenient for the study of the magnetic behaviour of ferromagnetic materials with different magnetic phases and/or dynamic properties. Particularly, the out-phase component gives useful information about the energy losses which ocurr during the domain wall movement and domain magnetization rotation by the *AC* magnetic field *h*. For instance, the study of both the real and imaginary parts of the $\chi_{ac}(T)$ has revealed the dynamics of domain walls in RAl$_2$ (R = Dy and Er) compounds [17]. It is worth noticing that most of the *AC* susceptibility results, including those reported for the milled GdX$_2$ Laves phase compounds, are constrained to the in-phase component.

In the present work a study of the ball milled cubic TbAl$_2$ Laves phase alloy has been undertaken. This compound crystallizes in the cubic MgCu$_2$ structure and orders ferromagnetically as other rare earth RAl$_2$ Laves phase compounds, with $T_C$ = 105 K [18]. The value of the coercivity and remanence due to the magnetocrystalline anisotropy is larger respect of the Gd compounds as suggested from the value of anisotropy constant reaching K = 7 x 10$^6$ J/m$^3$ [19]. Hence, in particular we will explore in more detail the *AC* susceptibility data to establish a comparison between the magnetic properties of bulk and milled TbAl$_2$. This will enable us to understand the effect due to the introduction of disorder. In addition, a deep structural characterization is also presented to correlate such effects with the eventual presence of single-domain magnetic grains. Finally, given the implications of the variations of the anisotropy with a possible grained structure at the nanoscale, we will also focus on the variation of the coercivity with grain size, which is a fingerprint of magnetic coupling/uncoupling processes among clusters/grains [20].

## 2. Experimental.

The starting alloy was prepared in an arc furnace from stoichiometric amounts of Tb (3N Alfa) and Al (5N, Alfa) metals. Then, a mass of 5 g of alloy was crashed and milled in a planetary high energy ball system Retsch PM 400/2 at a rotation speed of 200 rpm, using a container and balls made of tungsten carbide. Small amounts of the material were removed after 20, 70, 180 and 300 hours of milling to perform measurements, keeping always the ball-sample weight ratio to about 12. Handling and storage of the collected samples were carried out in a glove box under argon atmosphere in order to prevent the oxidation of the samples. The magnetic properties were collected in a Quantum Design PPMS in the temperature range 5-300 K and frequencies between 10 Hz and 10 kHz for the $\chi_{AC}$ ($h = 1$ Oe) and magnetic fields up to 9 T for the DC-magnetization M(H). The structural characterization was performed collecting X-ray patterns in a Philips PW 1710 diffractometer, with CuK$_\alpha$ radiation. The structural parameters were determined with a Rietveld powder profile programme [21] using silicon as a standard to correct for the instrumental broadening.

## 3. Results

*3.1- Unmilled TbAl$_2$*

The X-ray diffractogram of the unmilled sample (henceforth referred as *0 hours*) is presented in figure 1. This starting compound was found to crystallize in the cubic MgCu$_2$-type of structure with a lattice parameter $a = 7.8619(6)$ Å, in agreement with that already reported [22]. The sample is a single phase, as all reflections coming from the Laves phase *Fd-3m* crystallographic space group were identified and no spureous phases were found.

Concerning the magnetic properties, the zero field cooled (ZFC) and field cooled (FC) curves of DC magnetic susceptibility (M/H) at magnetic field of 500 Oe for bulk sample is presented in figure 2. It is clearly evident the existence of irreversibility. The temperature derivative of the FC magnetization is plotted in the inset. From this we can derive a precise value of Curie temperature $T_C = 105$ K.

Figure 3 displays both in-phase ($\chi'(T)$) and out-phase ($\chi''(T)$) components of the *AC* susceptibility obtained for the driving field of $h = 1$ Oe and for frequencies ranging from 100 Hz to 10 kHz. In $\chi'(T)$, a frequency-independent transition shows a $T_C$ value in agreement with the value estimated from the DC magnetization results. This confirms the existence of a ferromagnetic-paramagnetic transition at $T_C$. A broad anomaly in $\chi'(T)$ between 30 K and 60 K is smeared out with the increase of frequency. The curves depicted for the imaginary part $\chi''(T)$ reveal such a contribution in more detail. Three features can be observed: the ferromagnetic transition at $T_C = 105$ K, a broad anomaly around 80 K (more visible at the lowest frequency of 100 Hz), and a peak around 40 K which shifts to higher temperatures with the increase of the frequency, as indicated by arrows. A similar behaviour has already been described in other cubic ferromagnetic rare-earth Laves phase as $DyAl_2$ [17]. It is known that the imaginary component of the AC susceptibility is related to the absorption of energy by a ferromagnetic bulk material, which is caused by the domain wall excitations. Across the domain wall the spins change directions gradually due to the competition between magnetocrystalline and exchange energy. Thus, the clear rise in the peak of the $\chi''(T)$ could be explained by dynamic domain-wall movements as a consequence of a thermal activation process. This process has been well described by a simple two-level model for the elementary excitations from the high to low energy levels [17,23]. This approach leads to an Arrhenius relaxation law in which, if plotted the logarithm of the frequency of the *AC* magnetic field versus the inverse of the $\chi''(T)$ temperature maximum, the values for the relaxation time ($\tau_0$) and the activation energy ($E_a$) can be estimated. Thus, the results of the *AC* susceptibility for the *0 hours* sample follows the above model and gives $\tau_0 = 1.3 \times 10^{-11}$ s and $E_a = 803$ K (69 meV). If compare with the results obtained for the ferromagnetic compounds $DyAl_2$ ($\tau_o = 1.5 \times 10^{-8}$ s, $E_a = 40$ meV) [17], and $Sm_2Fe_{17}$ ($\tau_o = 1.7 \times 10^{-13}$ s, $E_a = 530$ meV) [23] the value for the activation energy $E_a$ is the order of that in $DyAl_2$, but the relaxation time is between the values reported for $DyAl_2$ and $Sm_2Fe_{17}$. The differences in $\tau_0$ and $E_a$ are related to the magnetocrystalline anisotropy or softness (hardness) of the ferromagnetic material and close to the attempt time of disordered magnets [24].

In the *0 hours* sample, as in any ferromagnetic material, the existence of an spontaneous magnetization and the formation of magnetic domains is expected. The thickness ($\delta$) of the domain walls is determined by the competition between the magnetocrystalline anisotropy and exchange interactions. The width ($\delta$) can be roughly estimated from the values of the exchange stiffness (A) and anisotropy constant (K). The parameter A can be calculated from the values of the Curie temperature, the lattice parameter *a* and the number of the nearest neighbours z according to [25]:

$$A = \frac{3k_B T_C}{za}$$

Where $k_B$ is the Boltzmann constant. In consequence, for the 180º domain walls in cubic materials:

$$\delta = \pi\sqrt{\frac{A}{K}} = \sqrt{\frac{3k_B T_c}{zaK}} \qquad (1)$$

And the domain wall energy per unit area:

$$\gamma = 2K\delta \qquad (2)$$

In TbAl$_2$ : T$_C$ = 105 K, $a$ = 7.8619 Å, $z$ = 4, and for the anisotropy constant K = 7 x 10$^6$ J/m$^3$ [19]. Substituting these values in the expressions (1) and (2), values of $\delta$ = 1.4 nm and $\gamma$ = 20 mJ/m$^2$ are obtained, below the value of $\delta$ = 40 nm for Fe and 5 nm for SmCo$_5$ [25]. The higher values of the anisotropy constant in SmCo$_5$ and TbAl$_2$ lead to narrow walls in these materials when compared to Fe, which presents a lower magnetocrystalline anisotropy K = 4.8x10$^4$ J/m$^3$ and a higher value of T$_C$ = 1043 K [25].

*3.2. Milled TbAl$_2$ sample*s.

X-ray diffraction patterns for the milled series are shown in figure 1. A progressive broadening and a reduction in the intensity of the peaks can be observed. This clearly indicates a decrease in the grain size [9-14], but it is also expected, although sometimes overlooked, that in metallic compounds a lattice strain is provoked by the milling. Although, a Williamson-Hall analysis [26] can be carried out to extract both the grain size (D) and the strain ($\epsilon$), it is most complete to perform a profile-fitting procedure of the diffraction peaks by the Rietveld method, which is not common in nanometric materials [27]. The parameterisation of the Thompson-Cox-Hastings function establishes the variation of the profile peak width as a function of the scattering angle using Gaussian and Lorentzian peak shapes [28]. These peak contributions include two terms; one corresponding to the line broadening due to the size (proportional to 1/cos$\theta$) and another term, related to the strain broadening, depending on tan$\theta$. The FULLPROF programme allows to extract the crystallographic parameters and the values of D and $\epsilon$ using a peak profile analysis which also takes into account the instrumental broadening [21].

The results of the Rietveld refinement of the X-ray diffraction patterns are presented for the alloy milled for *20 hours* in figure 4, as an example. The refinement was carried out using the *Fd-3m* space group, a Thompson-Cox-Hasting function for the peak profile, fixing the instrumental resolution function obtained for Si standard and with results for the Bragg errors R$_B$ ≈ 15, which are reasonable for the X-ray diffraction pattern refinement. In consequence, X-ray diffraction allows to derive well-defined values for the grain size and strain, as was reported in other studies [29, 30], sampling a large volume in comparison with transmission electron microscopy images. The grain size decreases with the milling time as depicted in figure 5. This result will have consequences on the evaluation of the magnetization and susceptibility curves. The smallest particle size, obtained for *300 hours* milled sample, is around 14 nm. Further milling does not decrease this value. The general tendency is similar to that found in other milled systems such as Fe-Cu-Ag [27] and Fe-Al [31], which are examples of heterogeneous magnetic systems. Figure 5 shows that the strain increases up to $\epsilon$ ≈ 1 %, due to the introduction of disorder with the grinding time, similar to what is observed in GdAl$_2$ [10] or Fe-Cu-Ag [32]. The strain also reaches a saturation for grinding times larger than 300 hours. The inset of figure 5 shows the relative decrease of the lattice parameter $\Delta a/a = 0.15\%$ in the *300 hours* respect to the *0 hours* alloy as was found in milled $GdAl_2$ compounds [10]. For this last series the effect of the creation of site defects (the Al can substitute on the rare earth sublattice) was suggested [10, 11], but the decrease of size of the particles (to the nanometer scale) is another source for lattice variations, as some of the grains can present important lattice distortions.

Changes of the Curie temperature in $GdAl_2$ milled series have been correlated with the variation of the lattice parameter and with disorder induced by mechanical milling [9-11]. These can be monitored from the ZFC and FC curves at H = 50 Oe, as shown in figure 6 for milled $TbAl_2$ alloys. The irreversibility of the ZFC and FC curves for the magnetic field of 50 Oe resembles the behaviour observed for reentrant spin-glasses. Nonetheless, the FC curves do not saturate and the irreversibility is high when compared to the canonical spin-glass. In this case the system of particles seems to behave as a cluster-glass. A decrease in the $T_C$ down to 102 K for the *20 hours* milled sample can be noticed in M/H curve and appears clearly defined in the derivative shown in the inset of figure 6. Regarding the 300 *hours* milled sample the $T_C$ is barely observed in the M/H curves, but is defined in the derivative shown in the inset, with $T_C$ = 98 K. In ZFC curve a broad transition, confirmed in the inset, appears at 45 K. The DC susceptibility data (M/H) in the paramagnetic region and for temperatures above 200 K is well-described by the Curie-Weiss law. The effective magnetic moment does not change significantly with $\mu_{eff}$ = 9.5 ± 0.2 $\mu_B$ along the series, near to the theoretical value for a free $Tb^{3+}$ ion of 9.72 $\mu_B$. The positive sign of the paramagnetic Curie temperature ($\theta$) indicates the presence of ferromagnetic correlations and the decrease from 106 K for the bulk to 7 K for 300 hours milled sample points towards a weakening of the exchange interactions [24]. The isothermal magnetization curves at T = 5 K as function of the magnetic field reveal a decrease in the saturation magnetization ($M_s$) when increasing the grinding time, as observed in figure 7. This evolution suggests a canting of the magnetic moments at high magnetic fields, which can be due to the strong disorder and random anisotropy present in the alloy, similar to that observed in amorphous FeZr [4]. In addition, the grinding process results in the presence of nanometric grains implying a canting of spins on the surface of the magnetic grains. Consequently, in this nanostructured $TbAl_2$, the domain wall movement mechanism for multidomain particles (observed in the bulk state) gives way to properties related to nanoparticle state of the alloy.

The value of the coercivity ($H_C$) is strongly influenced by the magnetocrystalline anisotropy in the ferromagnetic state and by the grain size. The value of $H_C$ = 68 Oe can be measured from the hysteresis loops, as shown for the *300 hours* milled alloy in the inset of figure 7. A decrease in $H_C$ with grinding time is observed and can be correlated to a decrease of the grain size. This decrease on going to smaller magnetic grains is related to the crossover from multidomain to single-domain magnetic grains [33]. The critical size for the appearance of single domain particles can be estimated from the expressions for magnetostatic and domain wall energies as follows [25]:

$$\frac{E_{ms}}{V} = \frac{\mu_0 M_s^2}{6}, \text{ and } V = \frac{4}{3}\pi r^3$$

Where $\mu_0$ is the permeability of vacuum, $M_s$ the saturation magnetization and $r$ the radius of the particle. Thus, the reduction of magnetostatic energy as result of the division of the particle into two domains is $\Delta E_{ms} = E_{ms}/2V$. The wall energy for the particle with area $\pi r^2$: $E_{wall} = \gamma \pi r^2$, being $\gamma$ the energy density per unit area. The reduction of energy by splitting into two domains will be:

$$\Delta E = \gamma \pi r^2 - \pi \frac{\mu_0 M_s^2 r^3}{9}$$

The condition for the critical size ($r_c$) is $\Delta E = 0$. Therefore, the critical radius of the single domain particles is given by:

$$r_c = \frac{9\gamma}{\mu_0 M_s^2}$$

From the M(H) curve as function of magnetic field at 5 K of the *0 hours* sample the value of $M_s$ = 8.52 $\mu_B$/mol is obtained. Taking $\rho$ = 5.82 g/cm$^3$ from the Rietveld refinement of the X-ray data, the conversion to the suitable units gives $M_s$ = 1.287x10$^6$ A/m. Using the value for the domain wall energy density estimated above, a critical radius for single domain particles of $r_c$ = 85 nm is calculated for the TbAl$_2$ alloy. Taking into account the average size obtained form the Rietveld refinements, <D> = 14 nm to 39 nm, we expect single-domain particles for all milled samples.

Additional information about the evolution of the magnetic phases can be furnished by the temperature dependence curves of $\chi'(T)$ and $\chi''(T)$, as shown in figure 8. The curves were obtained at $\nu$ = 10 Hz and $h$ = 1 Oe. The real component shows a smearing of the ferromagnetic transition and a peak around $T_f$ = 45 K for the *300 hours* milled sample. The peak related to $T_C$ shifts down to lower temperatures with the decrease of the particle size, whereas, the tendency is not clearly observed for $T_f$. The nature of the peak at $T_f$ = 45 K is further explored in the *300 hours* milled sample around the maxima. The $\chi_{ac}(T)$ results for this sample at different frequencies are presented in figure 9. It is particularly evident that the value of the maximum for $\chi'(T)$ (increases) and $\chi''(T)$ (decreases) with the increase of the frequency. This behaviour is standard in the magnetic relaxation of spin clustered systems. Furthermore, the value for the relative shift per frequency decade $\Delta T_f/T_f\Delta\log\nu$ = 0.0026(3) is similar to those reported for glassy magnets [24]. Quantitatively, the frequency dependence of the maximum can be reproduced by a phenomenological Vogel-Fulcher law using $\nu_0$ = 10$^{13}$ Hz, a value typical for spin-glass systems [24]. The result of the fitting yields a value of $T_0$ = 41 K for the interaction temperature and $E_a/k_B$ = 64 K for the activation energy, as shown in the inset of figure 9 a. Thus, a collective process of freezing of the spins rather than the domain wall excitation is observed, as single-domain particles (D< $r_C$) are present in the milled TbAl$_2$ structure.

A further step to analyse the spin-glass transition can be performed by studying the field dependence of M(T). In the ZFC curves for magnetic fields 1 kOe < H< 8 kOe depicted in figure 10, a shift of $T_f$ to lower temperatures with the increase of the DC magnetic field is observed. In the inset, the freezing temperature $T_f$ as function of the magnetic field is plotted. The $T_f$ (H) follows a $H^{2/3}$ law (*Almeida -Thouless line*), which gives another proof for the freezing (*collective*), as is commonly accepted [34].

## 4. Discussion

The results of the milling process in TbAl$_2$ polycrystalline alloy indicate, as seen in figure 1, a progressive decrease of the grain size and the absence of any new phase in the powder. The size reduction is commonly found in metallic and non-metallic systems, subjected to a mechanical milling process [16]. The trend for the crystalline size (D), and a concomitant increase of strain (ε) with milling time has been evaluated using Rietveld refinement. As commented in the former section, a minimum size of D = 14 nm and maximum ε ≈ 1% for the *300 hours* milled sample has been obtained. This values for size and strain are similar to other *cubic* Rare Earth-alloys or Transition metal [10, 32]. In addition, the reduction in size is also connected to the variation of lattice parameters (Δa/a), in coincidence with former studies in GdAl$_2$ [10].

The magnetic behaviour of the $\chi_{ac}$(T) for the bulk TbAl$_2$ alloy reflects the existence of a frequency-independent $T_C$ = 105 K, with an irreversible magnetic behaviour in the ZFC-FC curves. The ZFC curve is similar to that of $\chi'$(ν,T), although the lower *AC* driving field reflects more clearly the presence of strength of the anisotropy. This alloy is considered a ferromagnet [18] and the irreversibility stems from the competition between the anisotropy and magnetization process as a result of the presence of magnetic domains [17, 23]. Indeed, the variations in $\chi''$(ν, T) are similar to domain wall movements, already described in DyAl$_2$ and in agreement with the understanding of an energy loss process due to its absorption by the domain walls [17, 23]. Nevertheless, it is striking to note that such a behaviour is reminiscent of clustered magnetic (metallic) systems of crystalline and amorphous nature [35, 36], in which a local anisotropy is influencing the magnetic behaviour. Hence, it might be plausible that in this particular system the size of domains are reduced and be close to the nanometers scale, as it is the case in magnetically disordered amorphous FeZr-based alloys [5]. We will discuss in the following the variation produced by reducing the size of the particles.

In a fine-particle system the individual relaxation of the net magnetization of the magnetic grains gives rise to blocking/unblocking processes [7]. However, if the interparticle distances are small in those systems, strong interactions are expected among the particles. In our case, the structural study shows the existence of particles. These particles are below the single-domain size and, consequently, should show magnetic relaxation . However, the existence of grains in contact or with a narrow grain boundary modifies the relaxation phenomena leading to a behaviour which is sometimes labelled as *superferromagnetism* [37]. This is very dependent on the nature of the grain boundary. If this grain boundary is paramagnetic the magnetic atoms can become polarised and a ferromagnetic coupling can give rise to great modification in the coercivity field [20, 38]. If the magnetic grain boundary is ferromagnetic, the connection between the grains is obvious, and the behaviour would match that of the bulk alloy. A qualitative picture of the magnetic structure of mechanically milled $GdAl_2$, composed by paramagnetic $GdAl_2$ grains, ferromagnetically aligned Gd-rich clusters in the grain boundaries, and a Gd-Al grain boundary has been proposed by Williams D S et al. to explain the magnetic properties of this alloy [12].

By measuring the M(H,T) and $\chi_{ac}(T)$, we can extract parameters leading to further insight. On the one hand, the $T_f(H)$ obtained from $M_{DC}(H,T)$ follows a dependence ($H^{2/3}$) of the ZFC maxima which is associated to glassy magnetic behaviour [34]. Recent reports on ferromagnetic nanograins with a large degree of interparticle (dipolar) interactions, suggest that the variation should follow a $H^{1/2}$ [39]; however, this is not the case in the $TbAl_2$ nanoparticle samples. On the other hand, the $\chi_{ac}(T)$ presents a dynamic behaviour in which, although the general shape is not far from that of bulk-$TbAl_2$, the analysis of the $T_f(\nu)$ gives a low value for the relative shift per frequency decade of 0.0026(3) and can be fitted with the Vogel-Fulcher phenomenological law with $T_0 = 41$ K for the interaction temperature and $E_a/k_B = 64$ K for the activation energy  in the *300 hour*s-$TbAl_2$. This law is typically employed in magnetically disordered systems [24]. Thirdly, it is also worth mentioning here, that the 70 h and 180 h samples present a similar freezing temperature (approx. 45 K) whereas the size of grains is different (27 and 18 nm, respectively). This evidence seems to indicate that the freezing process is not strongly affected by the size reduction of particles. Overall, these three experimental observations seem to suggest initially that the alloys present an intrinsic glassy behaviour. Such a behaviour has been interpreted as a result of the modification of the lattice parameters in the grain structure, giving rise to a quadrupole site defect mechanism where the Al atoms can substitute on the rare earth sublattice, already proposed for the $GdX_2$ milled series [10, 11]. This supplies the disorder and frustration within the magnetic grains, giving rise to a random anisotropy. Nevertheless, in other bulk intermetallic alloys of $CeNi_{1-x}Cu_x$, a magnetic clustering with nanometric spin correlations is found due to the substitution of Ni and Cu at random, presenting behaviours for M(H, T) and $\chi_{ac}$(T) similar to those reported here [3]. Accordingly, the clustering/grain effects can not be ruled out in alloys in which the presence of nanometric grains is evidenced by X-ray diffraction data.

A clear indication that the nanostructure is in an ingredient of the magnetic behaviour is provided by the $H_C$(D). The variation of the coercivity with the dimension of the particles can be divided in two regions corresponding to multidomain and single magnetic domain. In the first region starting from the larger size, a reduction of the particle size leads to more pinning sites that result in the increase of $H_C$. Below the critical size the particles become single-domain and as the particle size decreases, $H_C$ falls down as $H_C = g - h/D^{3/2}$, where g and h are constants [33]. For the TbAl$_2$-milled samples we have obtained particles between 14 and 39 nm. These values are below the estimated critical radius of $r_c$ = 85 nm, where the particles become single-domain. Consequently, the coercivity should decrease with the grain size reduction as it is found here and shown in figure 11. The dependencies found in our case agree with a decrease of $H_C$ proportional to $D^{-3/2}$. This is striking as it suggests that the grain boundary, probably formed by very distorted nanometric grains, is very weak from the magnetic standpoint allowing a quasi-independent relaxation of the magnetic TbAl$_2$ grains. It also turns out that, if we compare the $H_C$(D) with those of ferromagnetic grains surrounded by a paramagnetic grain boundary, as observed in many ultrasoft nanocrystalline alloys of the Fe-Nb-Cu-Si-B and Fe-Zr-Cu-B types [37], our results appear consistently away from their predictions, with a $H_C$(D) decay proportional to $D^6$ due to a randomization of the anisotropy [20].

In view of the discussion just presented and the evident signs of collective freezing connected to a very disordered magnetic structure in the grains presented above, we propose that the magnetic structure of the nanostructured TbAl$_2$ is a collection of nanometric grains with sizes below the single domain limit, in which the magnetic structure is highly disordered due to the modification of the lattice parameter favouring the existence of a random local anisotropy. This anisotropy is enough to promote a freezing at 45 K, for all cases, but not enough to overcome the anisotropy associated to the nanometric grain structure. Given that the existence of nanoparticles in different matrices is usually related to *superparamagnetic* or *superferromagnetic* [37] behaviours, it is possible to consider that milled TbAl$_2$ is an example of a *super spin-glass*, which has been suggested in other systems such as microcrystalline goethite [40], ferrite nanoparticles [41] and nanogranular Al$_{49}$Fe$_{30}$Cu$_{21}$ [42].

## 5. Conclusion

The influence of size effects on the magnetic properties of TbAl$_2$ alloy has been studied in mechanically milled alloys. Bearing in mind work already reported in the GdX$_2$ series, we have explored in more detail the results of the AC susceptibility as well as the changes in the coercivity upon milling through M(H,T) measurements. Our Rietveld refinement of room-temperature of the X-ray spectra indicates the formation of a nanometric grains down to 14 nm in the *300 hours* alloy, with a strain reaching to 1 %. This reduction is clearly correlated to the variation of coercivity depending on grain size due to the presence of nanoparticles, which are situated below the crossover between the multi and single-domain behaviours.

The dynamics of the magnetic relaxation, the field and size dependence of $T_f$ suggests the presence of an intrinsic disordered magnetic state inside the particles which leads to a collective freezing, as occurs in *super spin-glasses*. The study of particles of sizes within the crossover region will lead to a deeper evaluation on how the local anisotropy in the grains is affected by the anisotropy effects typical of single-domain particles connected through a grain boundary.

**Acknowledgements**

This work was supported by the Direction of the Universities of the Ministry of Science and Education of Spain under contracts MAT 2003-06815, MAT 2005-04178-C04-02, SB2003-0102 and Juan de la Cierva contract.

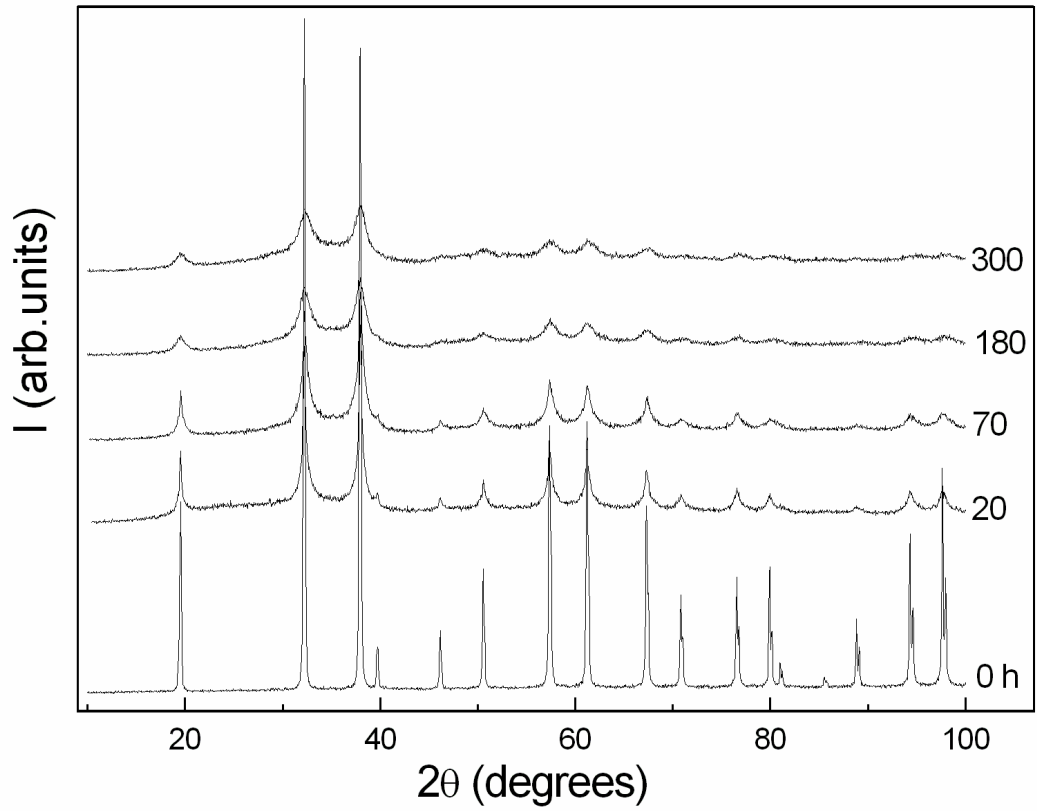

**Figure 1**. X-ray diffraction patterns for the as-cast (*0 hours*) and milled samples. The grinding time (in hours) is indicated on the right. Patterns have been shifted up for clarity. It is evident the large width of the peaks in the samples milled for long times.

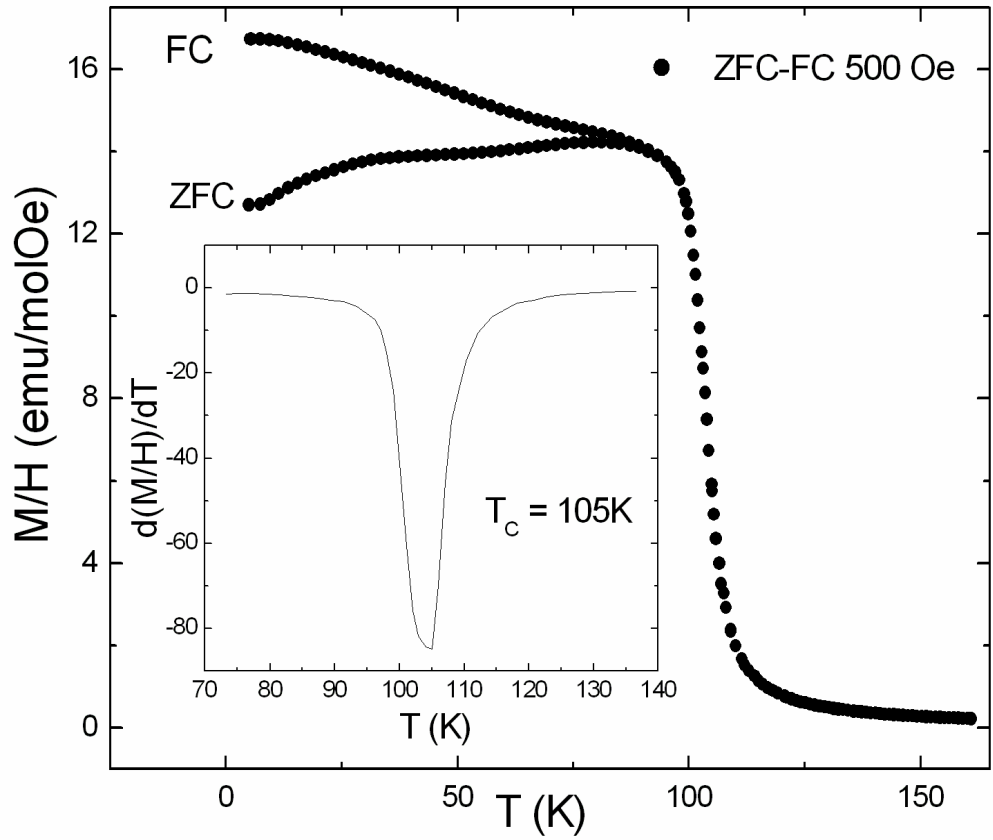

**Figure 2**. ZFC and FC curves of *DC* susceptibility at H = 500 Oe for the *0 hours* sample. The estimate of the transition temperature ($T_C$) is taken from the minimum of the first derivative, which is shown in the inset.

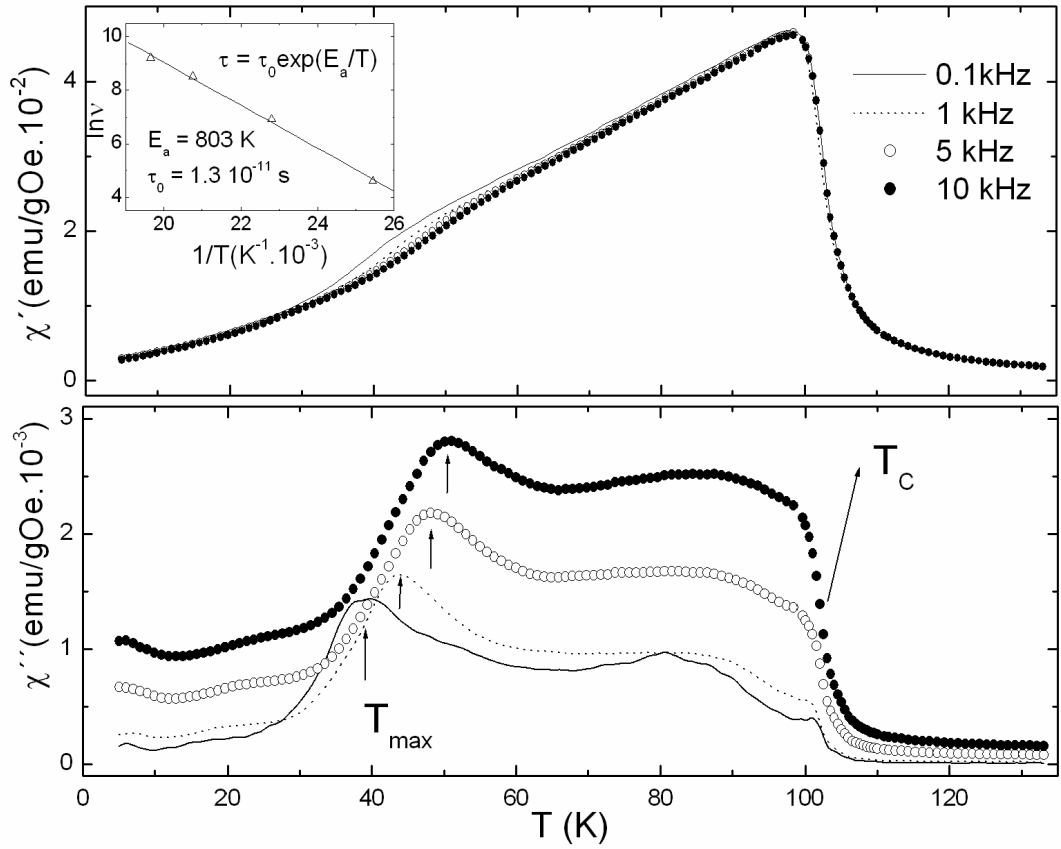

**Figure 3**. Thermal variation of the *AC*-susceptibility, $\chi'$ and $\chi''$, at several frequencies and $h = 1$ Oe for the TbAl$_2$ bulk sample. In $\chi''(T)$, the maximum around 40 K is associated, in magnetic bulk alloys, to domain wall motion. The dynamics of this process results in a shift of the maxima, which are indicated by arrows ($T_{max}$). This effect can be described by a simple thermal activation process (Arrhenius law), as shown in the inset.

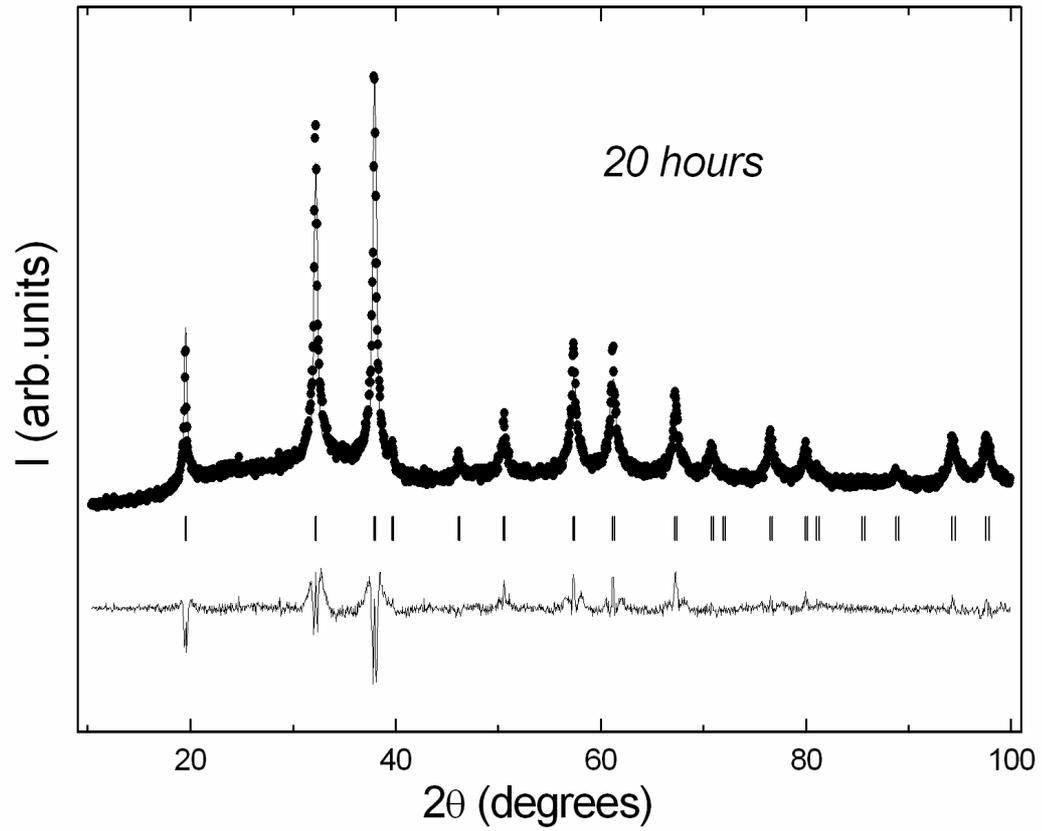

**Figure 4**. Rietveld refinement of the X-ray diffraction pattern for the *20 hours* milled sample. The experimental data are depicted by dots, the calculated refinement is represented by a continuous line and the difference is the bottom line. The Bragg positions are indicated by vertical markers.

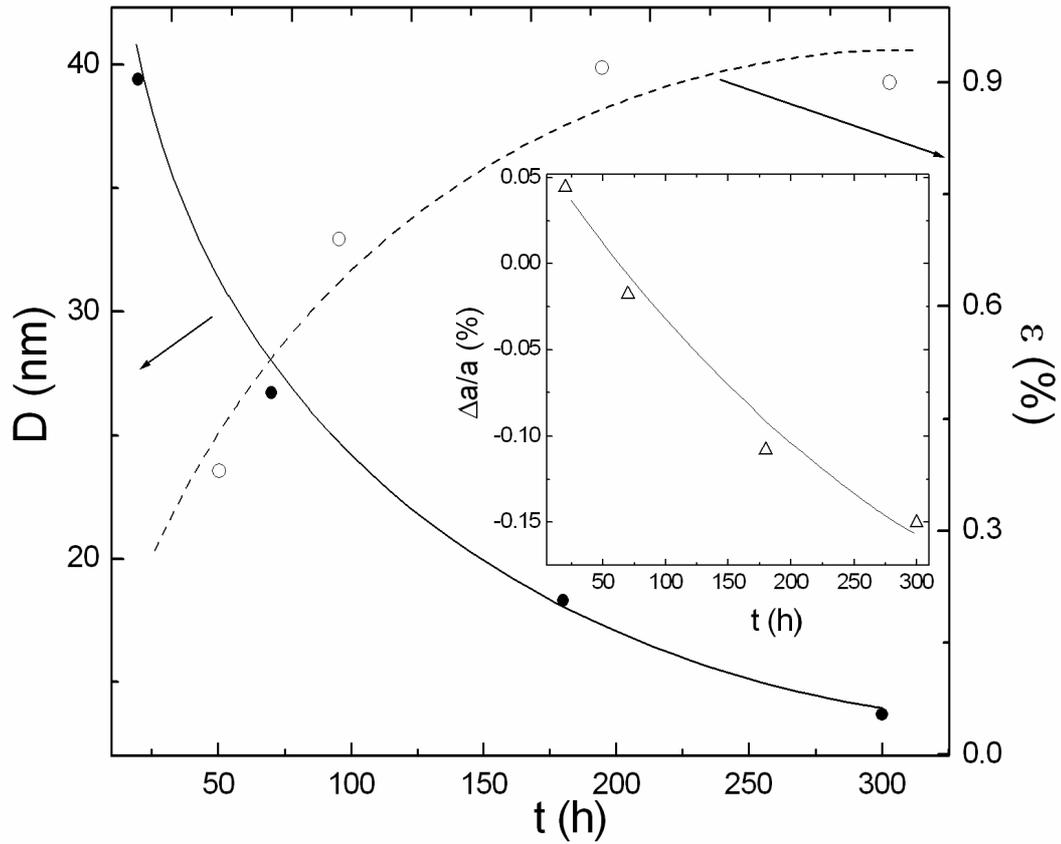

**Figure 5**. Particle diameter (D) (full symbols) and strain (ε) (open symbols) evolution as a function of the milling time. In the inset the relative change of the lattice parameter upon milling is presented. The lines are guides for the eyes. A diameter D < 20 nm is reached for *300 hours* of milling time.

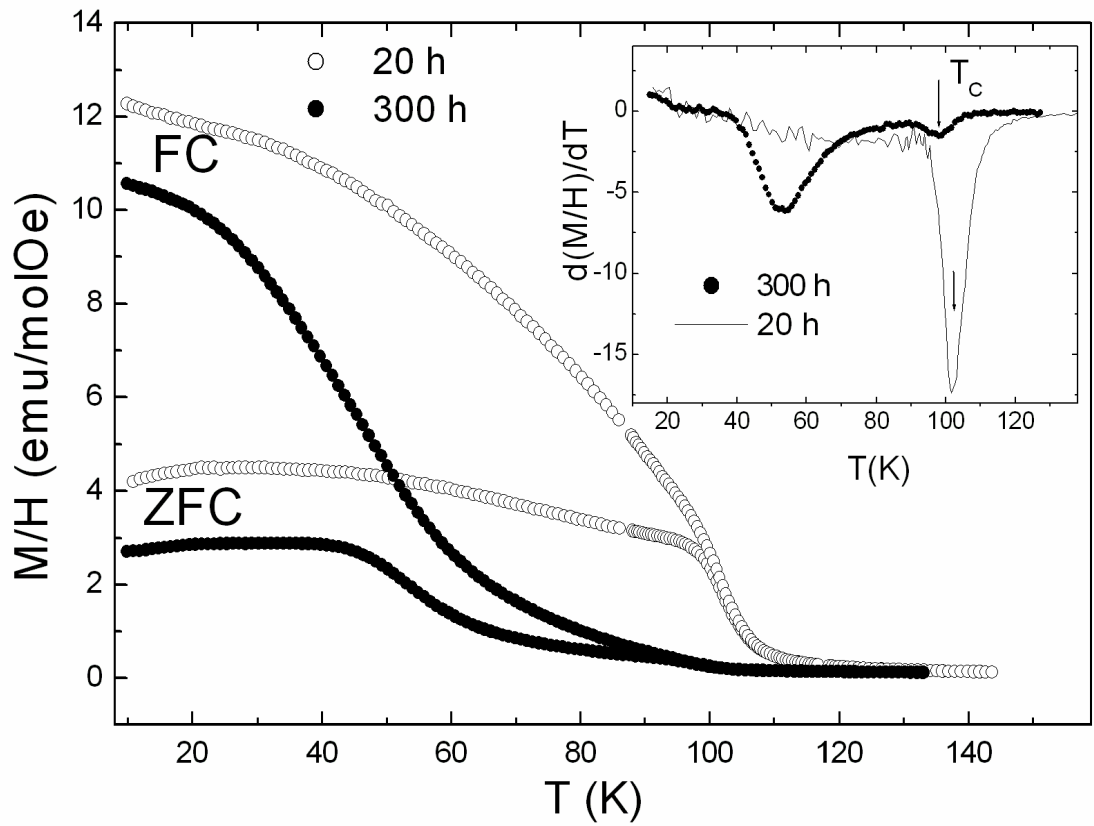

**Figure 6.** Temperature dependencies of ZFC and FC susceptibility at H = 50 Oe for the *20* and *300 hours* milled samples. A large irreversibility is found in both cases. $T_C$ is obtained from the derivative shown in the inset.

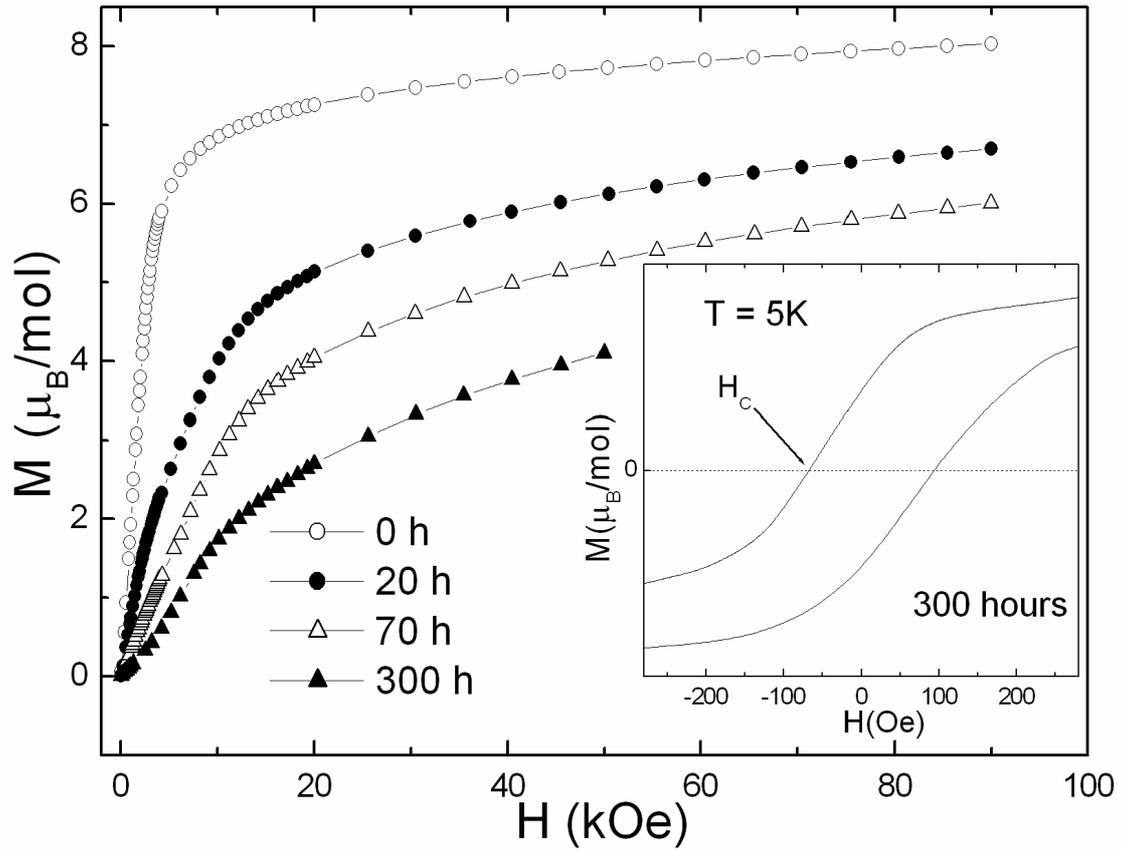

**Figure 7**. Magnetization curves as a function of the magnetic field for the *0 , 20, 70* and *300 hours* milled samples at 5 K. The low field region of the hysteresis loop for the *300 hours* milled alloy is shown in the inset. A value of $H_C \approx 70$ Oe is obtained for the coercivity.

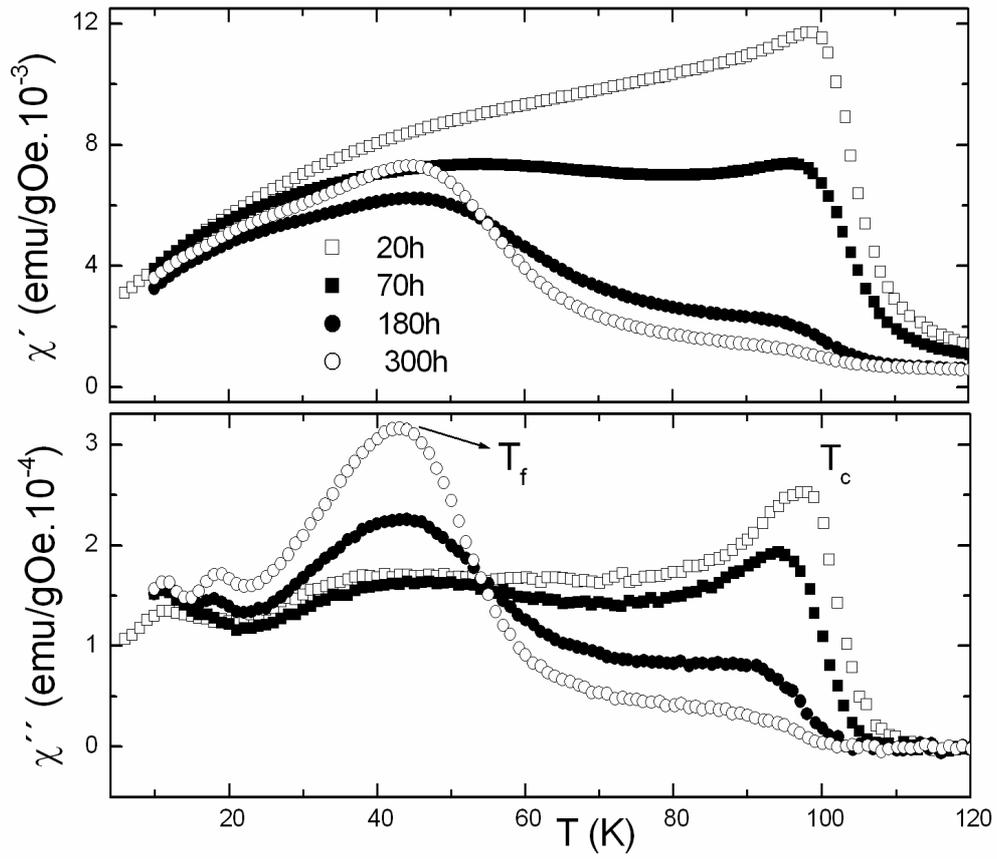

**Figure 8**. $\chi'(T)$ and $\chi''(T)$ for the series of milled samples at 10 Hz ($h$ = 1 Oe). The ferromagnetic contribution decreases with the increase of milling time, evolving to a glassy arrangement below the peak at $T_f$.

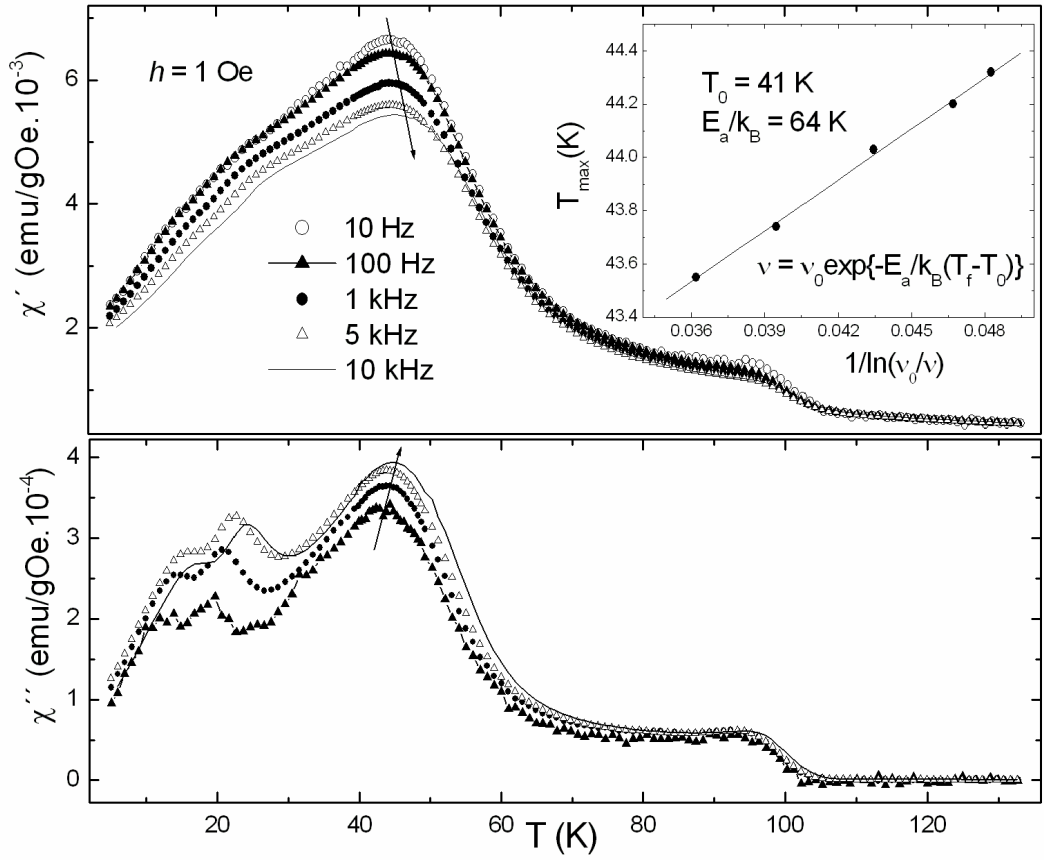

**Figure 9.** $\chi'$ (T) and $\chi''$ (T) at several frequencies for the *300 hours* milled alloy. In both cases, a shift of $T_f$ with the variation of frequency is observed. The inset shows a fit to the Vogel-Fulcher law for spin-glasses.

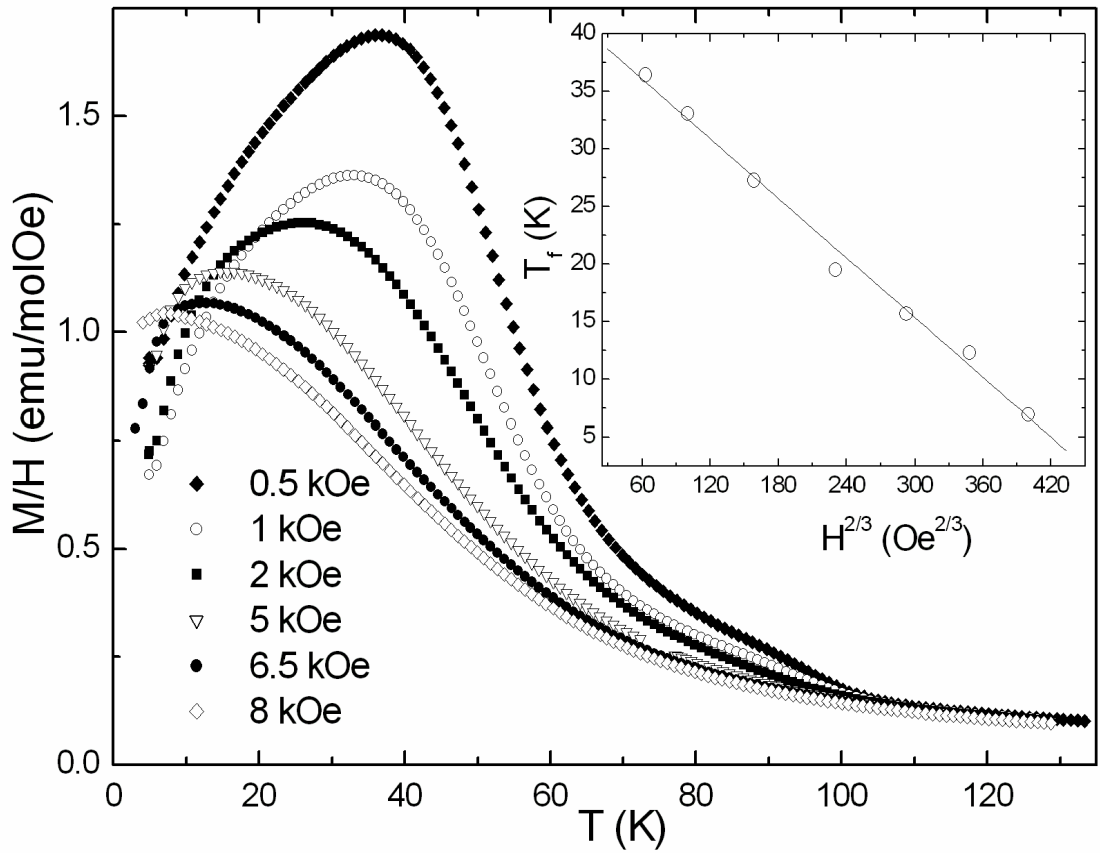

**Figure 10**. a) Magnetic field dependence of the ZFC susceptibility as a function of temperature for the *300 hours* milled sample. In the inset, the magnetic field dependence of the ZFC susceptibility maxima ($T_f$) follows a $H^{2/3}$ law (*Almeida-Thouless line*), as found in spin-glasses.

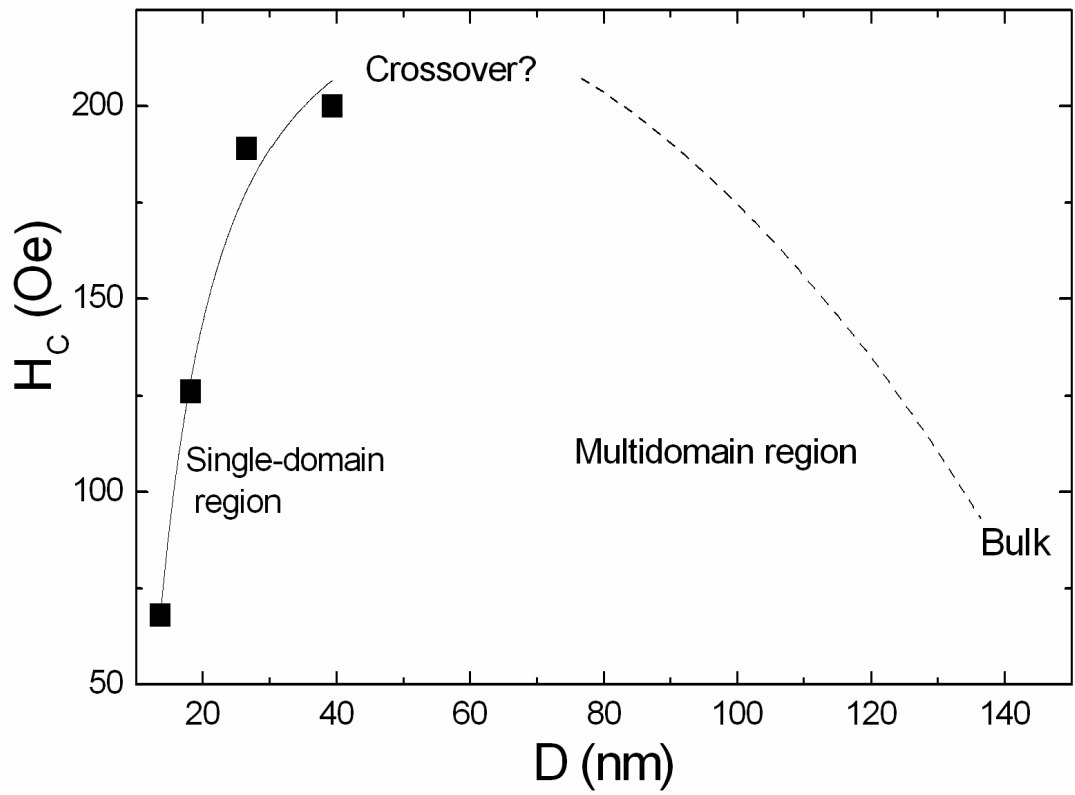

**Figure 11**. Size dependence of the coercivity ($H_C$). The milled alloys at *20, 70, 180* and *300 hours* appear within the single domain region. The dashed line marks the tendency between the bulk and nanometer size alloys. A crossover between both behaviours is expected around 60 nm. The solid line is the result of fitting to $H_C = g - h/D^{-3/2}$.